\documentclass[showpacs,amssymb,preprintnumbers,nofootinbib,superscriptaddress]{revtex4}

\usepackage{amsmath}
\usepackage{graphicx}
\usepackage{latexsym}
\usepackage{amsfonts}
\usepackage{url,hyperref}
\usepackage{bm}
\usepackage{textcomp}
\usepackage{color}

\newcommand{\nn}{\nonumber}
\newcommand{\beq}{\begin{equation}}
\newcommand{\eeq}{\end{equation}}
\def\bea{\begin{eqnarray}}
\def\eea{\end{eqnarray}}

\begin{document}

\title{Asymptotic iteration method for spheroidal harmonics of \\ 
higher-dimensional Kerr-(A)dS black holes}
\author{H.~T.~Cho}
\email[Email: ]{htcho@mail.tku.edu.tw}
\affiliation{Department of Physics, Tamkang University, Tamsui, Taipei, Taiwan, Republic of China}
\author{A.~S.~Cornell}
\email[Email: ]{alan.cornell@wits.ac.za}
\affiliation{National Institute for Theoretical Physics; School of Physics, University of the Witwatersrand, Wits 2050, South Africa}
\author{Jason~Doukas}
\email[Email: ]{j.doukas@ms.unimelb.edu.au}
\affiliation{Department of Mathematics and Statistics, The University of Melbourne, Parkville, Victoria 3010, Australia.}
\author{Wade~Naylor}
\email[Email: ]{naylor@se.ritsumei.ac.jp}
\affiliation{Department of Physics, Ritsumeikan University, Kusatsu, Shiga 525-8577, Japan}

\begin{abstract}
In this work we calculate the angular eigenvalues of the $(n+4)$-dimensional {\it simply} rotating Kerr-(A)dS spheroidal harmonics using the Asymptotic Iteration Method (AIM). We make some comparisons between this method and that of the Continued Fraction Method (CFM) and use the latter to check our results. We also present analytic expressions for the small rotation limit up to $O( c^3)$ with the coefficient of each power up to $O(\alpha^2)$, where $c=a\omega$ and $\alpha=a^2 \Lambda$ ($a$ is the angular velocity, $\omega$ the frequency and  $\Lambda$  the cosmological constant).
\end{abstract}

\date{12$^{th}$ April, 2009}
\maketitle

%%%%%%%%%%%%%%%%%%%%%%%%%%%%%%%%%
%
% Section 1: Introduction
%

\section{Introduction}\label{sec:intro}

\par Recently a new method for obtaining solutions of second order ordinary differential equations (with bound potentials) has been developed called the asymptotic iteration method (AIM) \cite{Ciftci:2003}. The AIM provides a simple approach to obtaining eigenvalues of bound state problems, even for spheroidal harmonics with $c$ a general complex number, large or small \cite{Barakat:2005,Barakat:2006}. It has also been shown that the AIM is closely related to the continued fractions method (CFM) \cite{Matamala:2007} derived from an exact solution to the Schr\"odinger equation via a WKB ansatz \cite{Miller}. A related CFM is often employed in numerical calculations of spheroidal eigenvalues and quasinormal modes of black hole equations \cite{Leaver:1985ax}, which is based on the series solution method of the Hydrogen molecule ion by Jaff\'e (and generalised by Baber and Hass\'e) \cite{Jaffe}. 

\par In this letter we will demonstrate that the AIM can also be applied to the generalized scalar hyper-spheroidal equation, $S_{kjm}(\theta)$, derived from an $(n+4)$-dimensional {\it simply} rotating Kerr-(A)dS angular separation equation \cite{Gibbons.G&&2005,Kodama:2008rq}:
\bea
 {1\over\sin\theta \cos^n\theta  } \partial _\theta \Big((1+\alpha \cos^2\theta) 
\sin\theta \cos^n\theta \partial_\theta S\Big)
+\left(A_{kjm}
    -\frac{m^2(1+\alpha)}{\sin^2\theta}  
       -\frac{c^2\sin^2\theta}{1+\alpha\cos^2\theta} 
    -\frac{j(j+n-1)}{\cos^2\theta} \right) S&&=0 \,\,\, , 
\label{angular}
\eea
where we have defined $\alpha=a^2 \Lambda$ with $a$ the angular rotation parameter. Note that $\Lambda<0$ corresponds to an asymptotic anti-de Sitter space, whereas $\Lambda >0 $ corresponds to an asymptotic de Sitter space \cite{Kodama:2008rq}, and the frequency $\omega$ is contained in the dimensionless parameter $c=a\omega$.

\par Higher dimensional spheroids have already been discussed by Berti et al. \cite{Berti}, who use a 3-term continued fraction method to solve the angular eigenvalues; however, the generalized scalar hyper-spheroidal equation under investigation there contains four regular singular points\footnote{Unlike the asymptotically flat limit ($\Lambda=0$) which only has three regular singular points.}, which leads to a 4-term recurrence relation \cite{Giammatteo:2005vu}. The simplest brute force approach to deal with an $n$-term recurrence relation is to use $n$ Gaussian eliminations to reduce the problem to a tri-diagonal matrix form \cite{LeaverGE}, but this can often be very tedious in practice. 
 
\par Even in four-dimensions the Kerr-(A)dS case does not allow for a simple 3-term continued fraction relation, nevertheless an elegant method has been developed to deal with situations of this type. In this case the CFM can be implemented by first transforming the angular equation into the Heun form \cite{Suzuki:1998vy,Kodama:2008rq}. However, such elegant techniques can only be applied if there are exactly four regular singular points in the equation. In contrast the AIM has an even broader appeal in that it can be applied somewhat\footnote{Of course some manipulation is first required to put the equation into the AIM form.} independently of the singularity structure of the ordinary differential equation, and thus to a larger class of equations without much modification or effort. The fact that the AIM may be of use when there are more than four singular points, or when other methods become prohibitively difficult, are compelling reasons to investigate it further. 

\par In this paper we use the AIM to find the eigenvalues of the {\it simply} rotating ($n+4$) Kerr-(A)dS spheroidal harmonics, but because this equation has four regular singular points we will also take this opportunity to compare the convergence rate of the AIM to that of the CFM (after first transforming into the Heun form). This serves as a double check of our results. 

\par The paper is organised as follows: In section \ref{sec:AIM} we give an overview of the AIM method and put our equation into the AIM form, then in section \ref{sec:Heun} we first transform our equation into the Heun form and then describe the CFM method. In section \ref{sec:smallc} we analytically calculate the first three coefficients of the eigenvalue in the small $c$ expansion, before finishing with some concluding remarks and analysis.       

%%%%%%%%%%%%%%%%%%%%%%%%%%%%%%%%%
%
% Section 2: The Asymptotic Iteration Method
%

\section{The Asymptotic Iteration Method}\label{sec:AIM}

\par To write the angular equation in a form suitable for the AIM we substitute $x=\cos \theta$ and obtain: 
\bea
(1-x^2) (1 + \alpha x^2)S''(x)+\left( {n(1-x^2)-x^2 \over x}+\alpha x(n+2 - (n+3)x^2 )-x(1 + \alpha x^2)\right)S'(x) && \nn \\
+\left(A_{kjm}-\frac{c^2 (1-x^2)}{1+\alpha x^2}-\frac{m^2 (1 +  \alpha)}{1-x^2}-\frac{j (j + n-1)}{x^2}\right)S(x)&=&0 \,\,\, . 
\label{genspher}
\eea
Note that the separation constant $A_{kjm}$ above corresponds to a simple eigenvalue shift in the asymptotically flat cases studied thus far \cite{Berti}, as can be verified by setting $\alpha=0$ (compare to equation (3.3) of reference \cite{Berti}). However, if $\alpha\neq 0$ such a shift is not possible, because of the non-trival $1 + \alpha x^2$ factor in the denominator. This means that the $\alpha \to 0$ limit will agree with the asymptotically flat case only after an eigenvalue shift. Also, note that because the above equation is invariant under $m\to -m$ we shall consider only $m\geq 0$.

\par The AIM can be implemented by multiplying $S_{kjm}$ by the characteristic exponents (as in section \ref{sec:Heun} for Heun's method); however, we have found that the most suitable form (fastest converging) is obtained by multiplying the angular mode function by \cite{Barakat:2005,Barakat:2006}:
\beq
S_{kjm}(x)= (1-x^2)^{|m|\over2}%x^j (1+\alpha x^2)^{\pm  \frac{i c}{2\sqrt{\alpha}} }  
y_{kjm} (x) \,\,\,
\label{scaling}
\eeq
%which leads to:
%
%\bea
%(1-x^2)(1+\alpha x^2)y''(x)+\bigglb[\frac{-2 \left(m+\left((m+2) x^2-1\right) \alpha %+1\right) x^2-n \left(x^2-1\right) \left(\alpha 
%   x^2+1\right)}{x} \biggrb]
%y'(x)&&\\
%+\bigglb[
%\frac{\left(x^2-1\right) c^2}{\alpha  x^2+1}+A_{kjm}-m (n+m+1)-m \left((n+m+3) x^2+m\right) \alpha
%   -\frac{j (n+j-1)}{x^2}
%\biggrb]y(x)&=&0\nn \,\,\, . 
%\eea
which leads to a differential equation in the AIM form:
\beq
y'' = \lambda_0y'+ s_0y\,\,\, ,
\label{AIM}
\eeq
where (for Kerr-(A)dS):
\bea
\lambda_0 & =  &-{1\over (1-x^2)(1+\alpha x^2)}\bigglb[\frac{-2 \left(|m|+\left((|m|+2) x^2-1\right) \alpha +1\right) x^2-n \left(x^2-1\right) \left(\alpha 
   x^2+1\right)}{x}\biggrb] \,\,\, , \\
s_0 & = & - {1\over(1-x^2)(1+\alpha x^2)}\bigglb[\frac{\left(x^2-1\right) c^2}{\alpha  x^2+1}+A_{kjm}-|m| (n+|m|+1)-|m| \left((n+|m|+3) x^2+|m|\right) \alpha
   -\frac{j (n+j-1)}{x^2}\biggrb] \nn\,\,\, , 
\eea
and where the primes of $y$ denote derivatives with respect to $x$. Differentiating equation (\ref{AIM}) $p$ times with respect to $x$, leads to:
\beq
y^{(p+2)} = \lambda_p y' + s_p y \,\,\, , 
\eeq
where the superscript $p$ indicates the $p$-th derivative with respect to $x$ and
\beq
\lambda_p = \lambda'_{p-1} + s_{p-1} + \lambda_0 \lambda_{p-1} \hspace{1cm} \mathrm{and} \hspace{1cm} s_p = s'_{p-1} + s_0 \lambda_{p-1} \,\,\, . 
\eeq
For sufficiently large $p$ the asymptotic aspect of the ``method" is introduced, that is:
\beq
\frac{s_p (x)}{\lambda_p (x)} = \frac{s_{p-1}(x)}{\lambda_{p-1}(x)} \equiv \beta(x) \,\,\, , 
\eeq
which leads to the general eigenfunction solution \cite{Ciftci:2003}:
\beq
y(x) = \exp \left[ - \int^x \beta (x') d\,x' \right] \left( C_2 + C_1 \int^x \exp \left\{ \int^{x'} \left[ \lambda_0 (x'') + 2 \beta ( x'') \right] d\,x'' \right\} d\,x' \right) \,\,\, , 
\eeq
for given integration constants $C_1$ and $C_2$, which can be determined by imposing a normalisation condition. Within the framework of the AIM, a sufficient condition for imposing termination of the iterations is when $\delta_p(x) = 0$, for a given choice of $x$, where  \cite{Barakat:2006}
\beq
\delta_p(x) = s_p(x) \lambda_{p-1}(x) - s_{p-1}(x) \lambda_p(x) \,\,\, .
\eeq
For each value of $m$ and $k$ (or $j$), in a given $(n+4)$-dimensions, the roots of $\delta_p$ leads a tower of eigenvalues $(m,\ell_1,\ell_2,\dots)$, where larger iterations give more roots and better convergence for higher $\ell$ modes in the tower. 

\par
\begin{figure}[h]
\centering
\scalebox{0.8}{\includegraphics{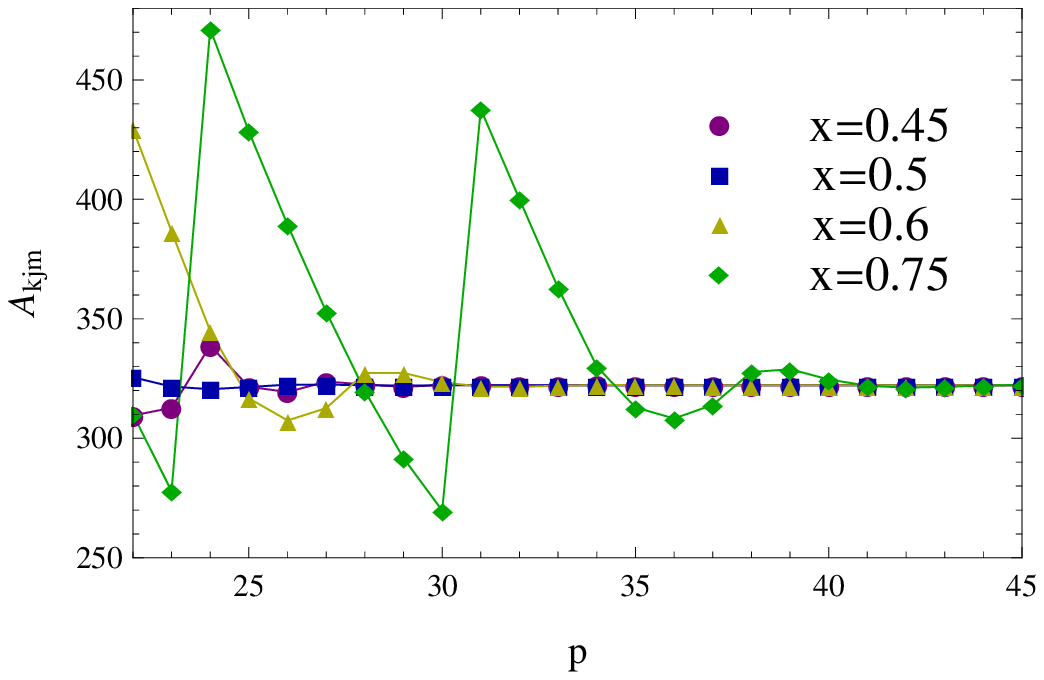}}
\hspace{0.5cm}
\scalebox{0.8}{\includegraphics{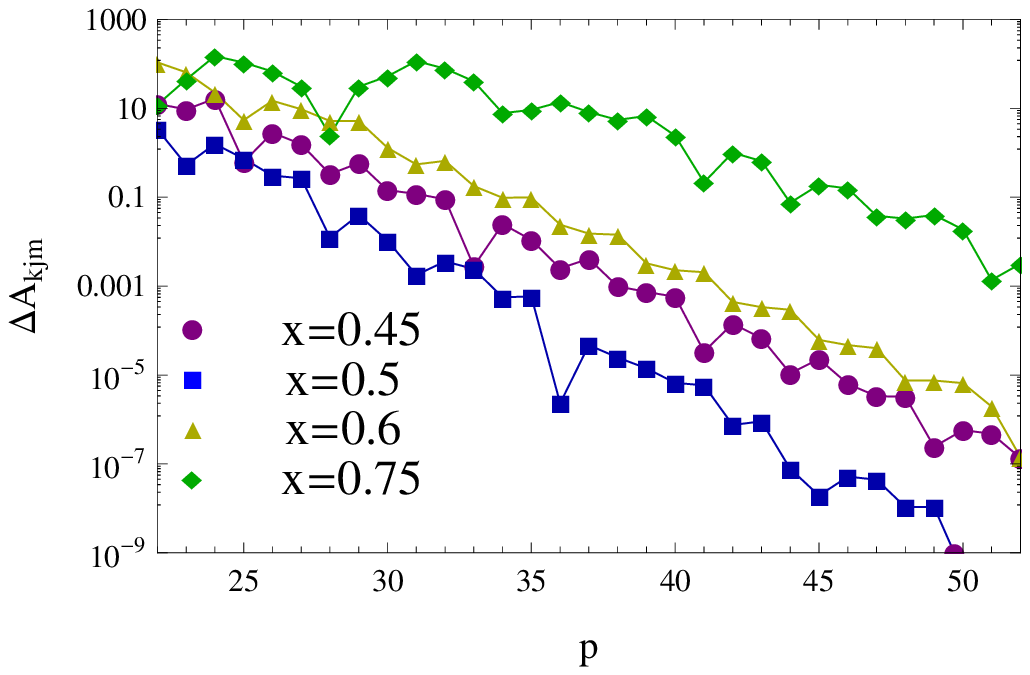}}
\caption{(Color Online) Plot of the convergence of a typical eigenvalue $A_{711}$ ($n=1$, $c=1$ and $\alpha=1$) under $p$ iterations of the AIM for various choices of $x=\{0.45,0.5,0.6,0.75\}$. Shown on the left is the eigenvalue versus $p$, while on the right is a $\log$ plot of the estimated error, $|A_{kjm}(p)-A_{kjm}(\infty)|$.}
\label{fig:aimcvergence}
\end{figure}

 It was noticed \cite{Barakat:2005,Barakat:2006} that the AIM converges fastest at the maximum of the potential, which in four dimensions occurs at $x=0$ (even with $\alpha \neq 0$ and for general spin-$s$). However, in the higher dimensional case we could not determine the relevant Schr\"{o}dinger like form and thus the maximum of the potential could not be analytically obtained. Nevertheless, as can be seen from the plots in Fig. \ref{fig:aimcvergence} we found that the point $x=\frac 1 2=\cos {\pi\over 3}$, in general, gave the fastest convergence. 

%%%%%%%%%%%%%%%%%%%%%%%%%%%%%%%%%
%%%%%%%%%%%%%%%%%%%%%%%%%%%%%%%%%%
\begin{figure}[h]
%\centering
\scalebox{0.81}{\includegraphics{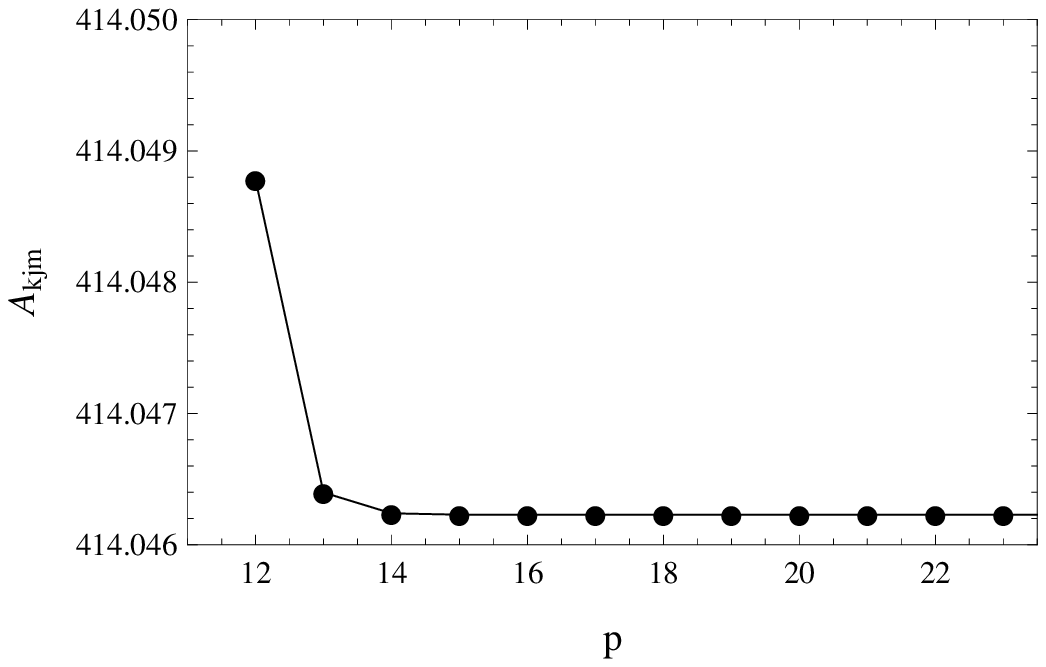}}
\hspace{0.5cm}
\scalebox{0.775}{\includegraphics{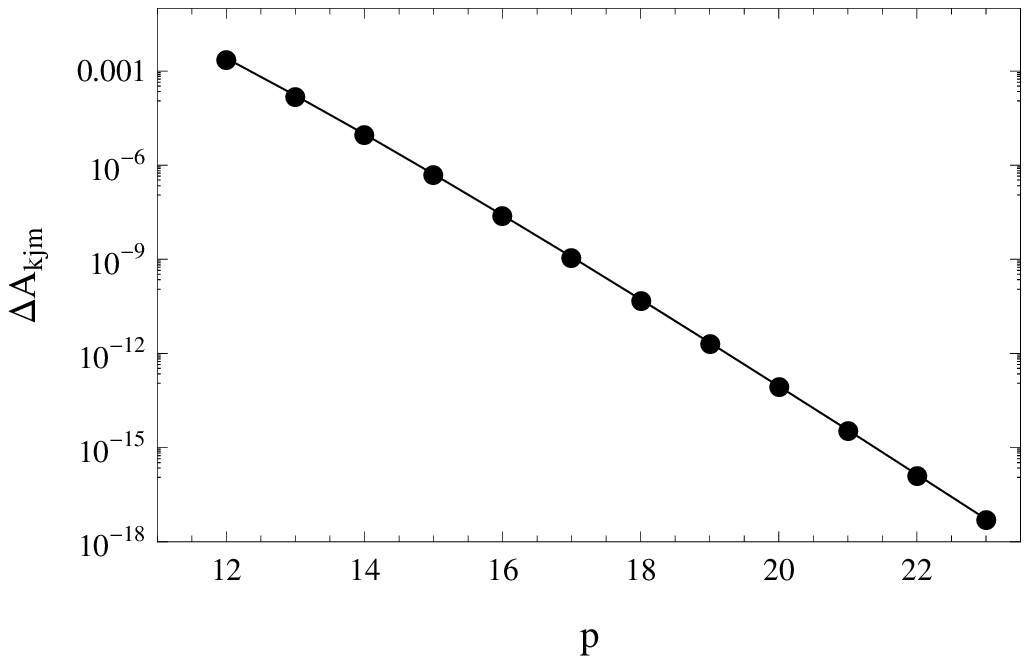}}
\caption{(Color Online) Plot of the convergence of a typical eigenvalue ($j=m=1$, $k=7$, n=1) under $p$ iterations of the CFM. The eigenvalue is shown on the left, while on the right is a $\log$ plot of the estimated error, $|A_{kjm}(p)-A_{kjm}(\infty)|$.}
\label{fig:CFMcvergence}
\end{figure}
%%%%%%%%%%%%%%%%%%%%%%%%%%%%%%

\par  One can also compare the relative rate of convergence between the AIM and CFM (see next section) methods for a typical eigenvalue by looking at the Log plots of the error, as shown in Fig.\ref{fig:aimcvergence} and Fig. \ref{fig:CFMcvergence}. Due to the typically exponential decrease in the error as a function of iteration number, if one is prepared to continue going to higher orders in the iterations of the AIM, eventually the AIM eigenvalues will exactly equal those of the CFM to any level of precision.

\par For the purposes of consistency we calculate each eigenvalue to 10 significant figures and put in brackets the minimum number of iterations required to reach this precision for both methods. From Tables \ref{tab:AdS1} - \ref{tab:DS3} it can be seen that larger $k$ modes require more iterations to achieve the same level of precision. Care should be taken when interpreting what information the iteration number gives us. Firstly, as one iteration of the AIM is not equivalent to one iteration of the CFM we can not relate this directly to computational speed. In fact although typically we need to iterate the AIM about twice as many times as the CFM the actual process time is often longer by factors of up to 100. That said we have not attempted to optimize either of the two methods here and it is not our current intention or purpose to do so. 

\par The results of the AIM for different values of $m$, $j$ and $k$ are presented in Tables \ref{tab:AdS1} - \ref{tab:DS3}. These results shall be discussed in section \ref{sec:conclusion} after discussing alternate approaches to obtaining the angular eigenvalue for Kerr-(A)dS spheroids.

%%%%%%%%%%%%%%%%%%%%%%%%%%%%%%%%%%%%%%%
\begin{table}[h]
\caption{\sl Selected eigenvalues, $A_{kjm}$, obtained from the AIM for the Kerr-AdS case with $c=1$, $\alpha=-0.05$, $n=1$ (extra dimensions) and $m=0$. Numbers in brackets represent the number of iterations required to reach convergence at the quoted precision, where subscript $A$ and $C$ are shorthand for AIM and CFM respectively. }
\label{tab:AdS1}
\begin{ruledtabular}
\begin{tabular}{llll}
$k$ &  $j=0$& $j=1$& $j=2$ \\
\hline
\hline
0 &   0.4978643318 (14)$_{A}$ (3)$_C$ &  3.317784170 (14)$_{A}$ (3)$_C$ & 8.201525517 (14)$_{A}$ (3)$_C$ \\
1 &    8.304871188 (15)$_{A}$ (4)$_C$ & 15.12466814 (15)$_{A}$ (4)$_C$ & 23.93276170 (16)$_{A}$ (4)$_C$ \\
2 &    23.89847347 (16)$_{A}$ (5)$_C$ &  34.63440913 (17)$_{A}$ (5)$_C$ & 47.35999084 (18)$_{A}$ (5)$_C$ \\
3 &    47.29227791 (17)$_{A}$ (6)$_C$ & 61.93248179 (19)$_{A}$ (6)$_C$ & 78.56537326 (21)$_{A}$ (7)$_C$ \\
4 &    78.48442957 (20)$_{A}$ (8)$_C$ & 97.02600365 (21)$_{A}$ (8)$_C$ & 117.5622674 (22)$_{A}$ (8)$_C$ \\
5 &    117.4747381 (23)$_{A}$ (9)$_C$ & 139.9165956 (23)$_{A}$ (9)$_C$ & 164.3544396 (24)$_{A}$ (9)$_C$\\
6 &    164.2631560 (25)$_{A}$ (10)$_C$ & 190.6047952 (24)$_{A}$ (10)$_C$ & 218.9432931 (27)$_{A}$ (10)$_C$ \\
7 &    218.8496664 (27)$_{A}$ (11)$_C$ & 249.0908236 (27)$_{A}$ (11)$_C$ & 281.3294495 (29)$_{A}$ (11)$_C$ \\
\end{tabular}
\end{ruledtabular}
\end{table}
 %%%%%%%%%%%%%%%%%%%%%%%%%%%%%%%%%
\begin{table}[h]
\caption{\sl Selected eigenvalues, $A_{kjm}$, obtained from the AIM for the Kerr-AdS case with $c=1$, $\alpha=-0.05$, $n=1$ (extra dimensions) and $m=j$. Numbers in brackets represent the number of iterations required to reach convergence at the quoted precision, where subscript $A$ and $C$ are shorthand for AIM and CFM respectively.}
\label{tab:AdS2}
\begin{ruledtabular}
\begin{tabular}{llll}
$k$ &  $j=m=1$& $j=m=2$& $j=m=3$ \\
\hline
\hline
0  &            8.302051467 (13)$_{A}$ (2)$_C$ &  23.90067165 (13)$_{A}$ (2)$_C$ &  47.29715054 (13)$_{A}$ (2)$_C$ \\
1 &              23.89902319 (15)$_{A}$ (3)$_C$ &  47.29444354 (15)$_{A}$ (3)$_C$ &  78.48871605 (16)$_{A}$ (4)$_C$ \\
2 &              47.29281932  (16)$_{A}$ (4)$_C$ &  78.48633467 (18)$_{A}$ (5)$_C$ &  117.4786244 (18)$_{A}$ (5)$_C$ \\
3 &              78.48490585 (19)$_{A}$ (6)$_C$ & 117.4764654 (15)$_{A}$ (6)$_C$ &  164.2667830 (20)$_{A}$ (6)$_C$ \\
4 &             117.4751699 (20)$_{A}$ (7)$_C$ & 164.2647680 (22)$_{A}$ (7)$_C$ &  218.8531204 (22)$_{A}$ (7)$_C$ \\
5 &             164.2635590 (23)$_{A}$ (8)$_C$ & 218.8512015 (24)$_{A}$ (8)$_C$ &  281.2375960 (25)$_{A}$ (8)$_C$\\
6 &             218.8500502 (26)$_{A}$ (9)$_C$ & 281.2357438 (26)$_{A}$ (9)$_C$ &  351.4201872 (27)$_{A}$ (9)$_C$ \\
7 &             281.2346325 (28)$_{A}$ (10)$_C$ & 351.4183829 (29)$_{A}$ (10)$_C$ &  429.4008807 (30)$_{A}$ (10)$_C$ \\
\end{tabular}
\end{ruledtabular}
\end{table}
 %%%%%%%%%%%%%%%%%%%%%%%%%%%%%%%%%
\begin{table}[h]
\caption{\sl Selected eigenvalues, $A_{kjm}$, obtained from the AIM for the Kerr-AdS case with $c=1$, $\alpha=1$, $n=1$ (extra dimensions) and $m=0$. Numbers in brackets represent the number of iterations, required to reach convergence at the quoted precision, where subscript $A$ and $C$ are shorthand for AIM and CFM respectively.}
\label{tab:DS1}
\begin{ruledtabular}
\begin{tabular}{llll}
$k$ &  $j=0$& $j=1$& $j=2$ \\
\hline
\hline
0 &	0.3796279195 (32)$_{A}$(7)$_C$ &	3.507876364 (29)$_{A}$(7)$_C$ &	
8.963027775 (32)$_{A}$(7)$_C$ \\ 
1 &	12.09034280 (29)$_{A}$(8)$_C$ &	21.46461733 (33)$_{A}$(8)$_C$ &	
32.77305405 (32)$_{A}$(8)$_C$ \\ 
2 &	34.99526905 (35)$_{A}$(9)$_C$ &	50.25395559 (36)$_{A}$(10)$_C$ &	
67.41776577 (35)$_{A}$(10)$_C$ \\ 
3 &	69.43716466 (39)$_{A}$(11)$_C$ &	90.47369703 (42)$_{A}$(11)$_C$ &	
113.4290146 (38)$_{A}$(11)$_C$ \\ 
4 &	115.3704526 (40)$_{A}$(12)$_C$ &	142.1641442 (42)$_{A}$(12)$_C$ &	
170.8870143 (41)$_{A}$(12)$_C$ \\ 
5 &	172.7901365 (44)$_{A}$(13)$_C$ &	205.3337886 (45)$_{A}$(14)$_C$ &	
239.8135617 (47)$_{A}$(14)$_C$ \\ 
6 &	241.6949846 (48)$_{A}$(15)$_C$ &	279.9853860 (48)$_{A}$(15)$_C$ &	
320.2166310 (50)$_{A}$(15)$_C$ \\ 
7 &	322.0845633 (51)$_{A}$(15)$_C$ &	366.1200572 (53)$_{A}$(15)$_C$ &	
412.0997175 (53)$_{A}$(15)$_C$ \\ 
\end{tabular}
\end{ruledtabular}
\end{table}
%%%%%%%%%%%%%%%%%%%%%%%%%%%%%%%%%
\begin{table}[h]
\caption{\sl Selected eigenvalues, $A_{kjm}$, obtained from the AIM for the Kerr-AdS case with $c=1$, $\alpha=1$, $n=1$ (extra dimensions) and $m=j$. Numbers in brackets represent the number of iterations, required to reach convergence at the quoted precision, where subscript $A$ and $C$ are shorthand for AIM and CFM respectively.}
\label{tab:DS2}
\begin{ruledtabular}
\begin{tabular}{llll}
$k$ &  $j=m=1$& $j=m=2$& $j=m=3$ \\
\hline
\hline
0 &	12.10323009 (26)$_{A}$(6)$_C$ & 	35.37074707 (24)$_{A}$(7)$_C$ &	
70.27222103 (25)$_{A}$(7)$_C$ \\ 
1 &	35.08856375 (31)$_{A}$(8)$_C$ &	        69.80790028 (31)$_{A}$(9)$_C$ &	
116.1867871 (28)$_{A}$(8)$_C$ \\ 
2 &	69.52978676 (36)$_{A}$(10)$_C$ &	115.7331075 (34)$_{A}$(10)$_C$ &
173.5944396 (32)$_{A}$(10)$_C$ \\ 
3 &	115.4610922 (35)$_{A}$(11)$_C$ &	173.1475066 (37)$_{A}$(11)$_C$ &
242.4916185 (37)$_{A}$(12)$_C$ \\ 
4 &	172.8794643 (38)$_{A}$(12)$_C$ &	242.0489748 (41)$_{A}$(13)$_C$ &
322.8761126 (38)$_{A}$(13)$_C$ \\ 
5 &	241.7834718 (45)$_{A}$(14)$_C$ &	322.4363105 (45)$_{A}$(14)$_C$ &
414.7467220 (45)$_{A}$(15)$_C$ \\ 
6 &	322.1724922 (50)$_{A}$(15)$_C$ &	414.3088831 (48)$_{A}$(16)$_C$ &
518.1027800 (50)$_{A}$(16)$_C$ \\ 
7 &	414.0462294 (53)$_{A}$(17)$_C$ &	517.6663481 (53)$_{A}$(17)$_C$ &
632.9438987 (53)$_{A}$(18)$_C$ \\ 
 \end{tabular}
\end{ruledtabular}
\end{table}
%%%%%%%%%%%%%%%%%%%%%%%%%%%%%%%%%
\begin{table}[h]
\caption{\sl Selected eigenvalues, $A_{kjm}$, via the AIM (we found that 32 iterations were required in every case to obtain the quoted precision) for different numbers of dimensions of the Kerr-dS case with $c=1$, $\alpha=1$, $m=j=k=0$.}
\label{tab:DS3}
\begin{ruledtabular}
\begin{tabular}{lllllll}
$n$ & 2 &3&4&5&6&7\\
\hline
$A_{000}$ & 0.2840487932 &  0.2253670617  & 0.1860599648 & 0.1580675679 & 0.1372036816 & 0.1210960725 \\ 
\end{tabular}
\end{ruledtabular}
\end{table}

%%%%%%%%%%%%%%%%%%%%%%%%%%%%%%%%%
%
% Section 3: Heun's method for the de-Sitter case
%

\section{Heun's method for de-Sitter case}\label{sec:Heun}

\par  As we mentioned earlier we could also work with a 4-term recurrence relation directly and use Gaussian elimination to obtain a 3-term recurrence, which then allows for the eigenvalues to be solved using the CFM. However, if we write the angular equation, equation (\ref{angular}), in terms of the variable $x=\cos(2\theta)$ \cite{Kodama:2008rq}, as: 
\bea
\label{angular-part}
(1-x^2) (2 + \tilde\alpha (1+x))S''(x)+\bigglb( n-1 -(n+3)x + {\tilde\alpha\over 2}(1+ x)(n+1 - (n+5)x) \biggrb)S'(x)&&\nn\\
+\Bigglb({A_{klm}\over 2}+\frac{c^2 (x-1)}{2(2+\tilde\alpha (1+x))}+\frac{m^2 (1 + \tilde\alpha)}{x-1}-\frac{j (j + n-1)}{x+1}\Bigglb)S(x)=0 \,\,\, 
\eea
and define $x=2z-1$, with the mode functions scaled by the characteristic exponents: 
\beq
Q(x)=  2^{|m|\over 2} (z-1)^{|m|\over 2}(2z)^{j\over2}\bigglb(z+{1\over \tilde\alpha}\biggrb)^{\pm {i c\over 2\sqrt{\tilde\alpha}} }y(z) \,\,\, ,
\eeq
{(note in this section we define $\alpha\to\tilde\alpha= a^2 \Lambda$ to avoid confusion with the standard Heun notation).}  The angular mode equation can now be written in the Heun form \cite{Suzuki:1998vy,Ron}:
\beq
\bigglb[\frac{d^2}{dz^2}+\left(\frac{\gamma}{z}+\frac{\delta}{z-1}+\frac{\epsilon}{z+{1\over\tilde\alpha}}\right)\frac{d}{dz}+\frac{\alpha\beta z-q}{z(z-1)(z+{1\over\tilde\alpha})}\biggrb]y(z)=0 \,\,\, , 
\eeq
where
\bea
&& \alpha =\frac{1}{2} (j + |m| \pm i{c\over \sqrt{\tilde\alpha}} )\,\,\, , \qquad\qquad
   \beta = \frac{1}{2} (j + |m| + n+3 \pm i{c\over \sqrt{\tilde\alpha}}) \,\,\, ,\\
&& \gamma=\frac{1}{2} (2 j + n+1) \,\,\, , \qquad\qquad
   \delta =1+|m| \,\,\,, \qquad\qquad
   \epsilon = 1\pm i{c\over \sqrt{\tilde\alpha}} \,\,\, ,\\
&\mathrm{and} & q=-\frac{m^2}{4}+ \frac{1}{4}(j\pm  i{c\over  \sqrt{\tilde\alpha}})(j+n+1\pm  i{c\over  \sqrt{\tilde\alpha}})- {1\over 4 \tilde\alpha} \Big[ (j+|m|)(j+|m|+n+1) -A_{kjm}\Big] \,\,\, , 
\eea
with the constraint 
\beq
\alpha+\beta+1=\gamma+\delta+\epsilon \,\,\, .
\eeq
Note that these results are identical to the Kerr-AdS case considered by Kodama et al. \cite{Kodama:2008rq} by choosing $\tilde\alpha = -a^2/R^2$ with $c=a\omega$.

\par To compare with the AIM method we shall use the fact that a three-term recurrence relation is guaranteed for any solution to Heun's differential equation \cite{Suzuki:1998vy,Ron}:
\bea
\alpha_0 c_{1}+\beta_0 c_0 &=& 0 \\
\alpha_p c_{p+1}+\beta_p c_p+\gamma_p c_{p-1}&=&0 \,,\qquad\qquad (p=1,2,\dots)\, ,
\eea
where for Kerr-(A)dS
\bea
\alpha_p&=&-
\frac{(p + 1)(p + r - \alpha + 1)(p+ r -\beta + 1)(p + \delta)}
{(2p + r + 2)(2p + r + 1)}\,\,\,,\\
\beta_p&=&\frac{\epsilon p(p+r)(\gamma-\delta)+[p(p+r)+\alpha\beta][2p(p+r)+\gamma(r-1)]}{(2p + r + 1)(2p + r - 1)}-{1\over\tilde\alpha}p(p+r)-q\,\,\,,\\
\gamma_p&=&-
\frac{(p + \alpha - 1)(p + \beta - 1)(p + \gamma - 1)(p + r - 1)}
{(2p + r - 2)(2p + r - 1)}\,\,\,,
\eea
with
\beq
r  = j+|m|+\frac{n+1}{2}\,\,\,.
\eeq
It may be worth mentioning that there is a removable singularity in $\beta_p$ for the initial condition $p=0$ with $r=1$ (when $n=1$ and $j=|m|=0$), which for the five-dimensional case, $n=1$, implies that  this initial condition must be treated separately. Once a 3-term recurrence relation is obtained the eigenvalue $A_{kjm}$ can be found (for a given $\omega$) by solving a continued fraction of the form \cite{Leaver:1985ax, Berti}:
\beq
\beta_0-\frac{\alpha_0\gamma_1}{\beta_1-}\frac{\alpha_1\gamma_2}{\beta_2-}\frac{\alpha_2\gamma_3}{\beta_3-}\ldots=0\,\,\,.
\label{CFM}
\eeq
We have used this method to compare with the AIM seen in Tables \ref{tab:AdS1} - \ref{tab:DS3}. The convergence of this method is shown for a typical representative eigenvalue in Fig. \ref{fig:CFMcvergence}, further results are discussed in section \ref{sec:conclusion}.

%%%%%%%%%%%%%%%%%%%%%%%%%%%%%%%%%
%
% Section 4: Analytic Results for Small Rotation
%

\section{Analytic Results for Small Rotation}\label{sec:smallc}

\par It is also useful to have some analytic expressions at hand for the angular eigenvalues. For small $c$, these can be obtained by standard perturbation theory \cite{Giammatteo:2005vu} or by using eigenfunction expansion methods \cite{Seidel:1988ue}. However, a very convenient approach well suited to symbolic computations is the method used by Berti et al. \cite{Berti}, also see reference \cite{Suzuki:1998vy}. In the limit $c\to 0$ the infinite series terminates at some finite $k$ and we are left with \cite{Kodama:2008rq}:
\beq
A_{kjm} = (2k+l+|m|)(2k+l+|m|+n+1) \,\,\, .
\eeq
Choosing $2k=l-(j+|m|)$, we find the correct $c=0$ limit: $A_{kjm}=l(l+n+1)$, with the constraint $l\geq j+|m|$. Then, in order to now find the small $c$ perturbative expansion of  $A_{kjm}$ it is convenient to use the inverted CFM, which is the $k^{th}$ inversion of equation (\ref{CFM}) \cite{Berti}:
\beq
\beta_k-\frac{\alpha_{k-1}\gamma_k}{\beta_{k-1}-}\frac{\alpha_{k-2}\gamma_{k-1}}{\beta_{k-2}-}\ldots\frac{\alpha_0\gamma_1}{\beta_0}=\frac{\alpha_k\gamma_{k+1}}{\beta_{k+1}-}\frac{\alpha_{k+1}\gamma_{k+2}}{\beta_{k+2}-}\ldots
\label{inv}
\eeq
and assume a power series expansion of the form: 
\beq
A_{kjm}=\sum_{p=0}^\infty f_p c^p\,\,\,.
\eeq
When we substitute this power series into equation (\ref{inv}) the terms $f_p$ can be found by equating powers of $c$ (after a series expansion of equation (\ref{inv})). For the asymptotically flat case \cite{Berti} it is very simple to go to large powers of $c$, where (in general) results to order $c^p$ can be obtained by going to order $k=p$ in equation (\ref{inv}). Unfortunately, for the Kerr-(A)dS case the expressions are complicated by the inclusion of the curvature term $\alpha= a^2 \Lambda $ (for example the four dimensional case is given in reference \cite{Suzuki:1998vy}). Although results for any value of $\alpha$ can be stored on a computer, they are too large to present on paper. Thus, given below are results up to and including ${\cal O} (c^2)$ with a further series expansion up to ${\cal O} (\alpha)$:  \bea
f_0 &=& l(l+n+1) \,\,\,,\\
f_1&=& {\alpha \over c\, (2l+n-1)(2l+n+3) } \bigglb(2 l^4+4 (n+1) l^3+(2 m^2+3 n (n+2)-1) l^2+(n+1) (2 m^2+n^2+2 n-3) l  \\
&&+m^2 (n+1) (n+3)-j^2 (n^2+2 l n+4 n+2 l (l+1)+3)-j (n-1) (n^2+2 l n+4 n+2 l (l+1)+3)\biggrb) \,\,\, , \nn \\
f_2 &=& -\frac{1}{c^2} \Big[ \frac{1}{2} \alpha  \Big(-\frac{c^2 \left(2 l^2+2 l (n-1)+n (n+2)+5\right) (j-l-|m|) (j-l+|m|)(j+l-|m|+n-1) (j+l+|m|+n-1)}{(2 l+n-3) (2 l+n-1)^3 (2 l+n+1)} \nn \\
&&\hspace{1cm}+\frac{c^2 \left(2 l^2+2 l(n+3)+(n+3)^2\right) (j-l-|m|-2) (j-l+|m|-2) (j+l-|m|+n+1) (j+l+|m|+n+1)}{(2 l+n+1) (2 l+n+3)^3 (2 l+n+5)} \nn \\
&&\hspace{1cm}+\frac{2}{(2 l+n-1) (2l+n+3)} \Big\{-2 j^2 \left(l^2+(l-1) n+l-3\right)-2 j l (n+3) (l+n+1)-2 l^4-4 l^3 (n+1) \nn \\
&&\hspace{1.5cm}+l^2 \left(2 m^2-3 n (n+2)+1\right)-l (n+1) \left(-2 m^2+n^2+2 n-3\right)-2 m^2 (n+3)\Big\} +2 j (j+n+1)-2 m^2\Big)\nn \\
&&\hspace{1cm}+c \left(f_1-\frac{2 c \left(-j^2-j n+j+l (l+n+1)+m^2+n-1\right)}{(2
   l+n-1) (2 l+n+3)}\right) \Big] \,\,\, .
\eea
The singular behaviour in the denominators of $f_1$ and $f_2$ always cancels for $n=1$, because the constraint $l\geq j+|m|$ with $2k=l-(j+|m|)$ being an integer, always leads to a zero in the numerator as well. This can be verified explicitly case by case. Note that if this expression is to be used for explicit numerical calculations then the limit must be chosen carefully.

\par It appears that in the case of Kerr-(A)dS, odd powers of $c$ also contribute to the spin-0 case (or  tensor part of the graviton perturbations). Note that in the limit $\alpha \to 0$ these results do not quite agree with the results given in reference \cite{Berti}, because of the eigenvalue shift in equation (\ref{genspher}). This can simply be remedied by adding $-c^2$ to the $f_2$ term. The exact eigenvalue solution is compared with the small $c$ expansion in Table \ref{tab:SmallC}, which shows good agreement. 

%%%%%%%%%%%%%%%%%
\begin{table}[h]
\caption{\sl Comparison of the small c expansion and the exact result (via the AIM or CFM) for the Kerr-(A)dS case with $j=1$, $k=m=0$ and $n=2$ (extra dimensions),  for given values of  $c$ and $\alpha$ (results quoted to $6 ~s.f.$).}
\label{tab:SmallC}
\begin{ruledtabular}
\begin{tabular}{ccccccc}
\hline
$(c,~\alpha)$ & (0.1,~0.05) &   (0.1,~-0.05)  &  (0.5,~0.1)  & (0.5,~-0.1)  & (0.5,~0.5)  & (0.5,~-0.5)  \\
\hline
$A_{010}$ &   4.01680 & 3.98833 & 4.09588 & 4.04457 & 4.17955 &3.90800 \\
\hline
${\rm Small}~c$  &   4.01708 & 3.98863 & 4.09700 & 4.04585 & 4.19930 & 3.94355 \\
\hline
\end{tabular}
\end{ruledtabular}
\end{table}

%%%%%%%%%%%%%%%%%%%%%%%%%%%%%%%%%
%
% Section 5: Conclusion
%

\section{Analysis \& Discussion}\label{sec:conclusion}

\par We have calculated to ten significant figures the eigenvalues shown in Tables \ref{tab:AdS1} to \ref{tab:DS2} of the $(n+4)$-dimensional {\it simply} rotating Kerr-(A)dS angular separation equation using the AIM. This generalises the results found in references \cite{Barakat:2005,Barakat:2006} to Kerr asymptotically de-Sitter or anti de-Sitter spacetimes. Our results were also checked using the CFM. Although we only considered a real parameter $c=a\omega$, we could also have used a purely imaginary or complex value of $c$, and thus, the AIM may be of use for quasinormal mode analysis. For brevity we presented results for $n=1$ extra dimensions only, but we have also checked the dependence on dimension, as can be seen in Table \ref{tab:DS3} for the fundamental $k=0$ mode. 

\par  All numerics and symbolic computations used MATHEMATICA$^{\textrm{\textregistered}}$ where we found that the CFM eigenvalue solutions converged very quickly with accurate results even after a continued fraction depth of only $p=15$. As a check of our numerics we also compared our CFM results with some independent CFM code \cite{Berti:2009kk}, where we found identical results. (Note, that because our method is symbolic we can use the {\tt NSolve} command in MATHEMATICA, as opposed to the method in reference \cite{Berti:2009kk} that uses {\tt FindRoot}). 

\par One point worth mentioning is that the $\alpha\to 0$ limit cannot be taken via Heun's method, because the recurrence relation (and hence the continued fraction) diverges for this case. In contrast the AIM has no such problem.  The AIM also gives an alternative approach to obtain the eigenfunctions in terms of simple integrals, which may be useful for symbolic computations. Another commonly used approach would be to the use the series solution method of Leaver \cite{Leaver:1985ax}.

\par We also obtained new results for the small $c$ expansion of the angular eigenvalue $A_{kjm}$ up to $O(c^3)$. Because of the complexity of these expressions, we only presented the the coefficient of each power up to $O(\alpha^2)$, where it is interesting to note that the small $\alpha$ expansion of the small $c$ series does allow us to obtain the $\alpha\to 0$ limit analytically from the CFM. 
{As future work it would also be interesting to compare this result with that obtained by standard perturbation theory \cite{Giammatteo:2005vu} or eigenfunction expansion \cite{Seidel:1988ue} approaches.}

\par It is also worth mentioning that the AIM (or Heun's method) can also be applied to the case when there are two or more rotation parameters ({\it non-simple}) \cite{Gibbons.G&&2005}, where some interesting studies have already investigated spheroids in five-dimensions for Kerr-AdS in the near degenerate (equal rotation) limit  \cite{Aliev:2008yk}. However, each dimension must be considered case by case, because a general expression for $(n+4)$ dimensions has not been found. We intend to report on the angular spheroids (and associated radial equations) for these interesting cases in the near future.

\par In conclusion, we have highlighted how the AIM can be applied to higher-dimensional scalar or tensor gravitational (for $n\geq 3$) spheroidal harmonics, which arise in the separation of metrics in general relativity. We have seen that the AIM requires very little manipulation in order to obtain a fast route to the angular eigenvalues, which may be useful for cases where Heun's method may not apply. However, the AIM does have some shortfalls, because although we did not attempt to optimise either algorithm, our implementation of the AIM was found to be much slower than that of the CFM. Considering that the CFM essentially involves expanding out $p$ nested fractions, whereas the AIM involves taking  $p^{\rm th}$ order derivatives, this behaviour is not surprising. However, for most of the cases we considered only a few seconds were required to reach the desired level of accuracy and thus, timing was not a large concern. 

\par We hope that the AIM might be of some topical use, for example, in the angular spheroids needed in the phenomenology of Hawking radiation from spinning higher-dimensional black holes \cite{Frost:2009cf}. We have recently used a combination of all the techniques discussed in this work to evaluate the angular eigenvalues, $A_{kjm}$, for real $c=a\omega$, which are needed for the tensor graviton emission rates on a {\it simply} rotating Kerr-de Sitter black hole background in $(n+4)$-dimensions \cite{Doukas:2009cx}.

%%%%%%%%%%%%%%%%%%%%%%%%%%%%%%%%%
%
% Acknowledgements
%

\acknowledgments

\par We are grateful to the authors of Reference \cite{Berti} for correspondence clarifying the non-singular behaviour of the small $c$ expansion and WN would also like to thank Max Giammatteo for useful discussions. HTC was supported in part by the National Science Council of the Republic of China under the Grant NSC 96-2112-M-032-006-MY3. Financial support from the Australian Research Council via its support for the Centre of Excellence for Mathematics and Statistics of complex systems is gratefully acknowledged by J.~Doukas.

%%%%%%%%%%%%%%%%%%%%%%%%%%%%%%%%%
%
% References
%

\end{document}